\documentclass[bib]{statapress}

\usepackage[crop,newcenter,frame]{pagedims}
\usepackage{sj,epsfig,stata,shadow}

\sjsetissue{$vv$}{$ii$}{$mm$}{$yyyy$}

\newcommand{\NormDist}{\mathrm{N}}
\newcommand{\UnifDist}{\mathrm{Unif}}
\usepackage{amsfonts}
\newcommand{\R}{\mathbb{R}}

\newcommand{\dconv}{\stackrel{d}{\to}}
\newcommand{\independenT}[2]{\mathrel{\rlap{$#1#2$}\mkern2mu{#1#2}}}
\newcommand\independent{\protect\mathpalette{\protect\independenT}{\perp}} 
\newcommand{\matf}[1]{\boldsymbol{\mathbf{\MakeUppercase{#1}}}} 
\newcommand{\vecf}[1]{\boldsymbol{\mathbf{\MakeLowercase{#1}}}} 
\usepackage{amsmath}
\DeclareMathOperator{\E}{E}
\DeclareMathOperator{\Q}{Q}
\DeclareMathOperator{\1}{\mathbf{1}}
\newcommand{\Ind}[1]{\1\{#1\}}
\newcommand{\diff}{\mathop{}\!\mathrm{d}} 
\let\Statanum\num
\let\num\relax
\usepackage{siunitx}

\let\num\Statanum

\begin{document}

\inserttype[st0001]{article}

\author{D.\ M.\ Kaplan}{%
  David M.\ Kaplan\\Department of Economics\\University of Missouri\\Columbia, MO\\kaplandm@missouri.edu
}

\title[Smoothed instrumental variables quantile regression]{Smoothed instrumental variables quantile regression}

\maketitle

\begin{abstract}
In this article, I introduce the \stcmd{sivqr} command, which estimates the coefficients of the instrumental variables (IV) quantile regression model introduced by \citet{ChernozhukovHansen2005}.
The \stcmd{sivqr} command offers several advantages over the existing \stcmd{ivqreg} and \stcmd{ivqreg2} commands for estimating this IV quantile regression model, which complements the alternative ``triangular model'' behind \stcmd{cqiv} and the ``local quantile treatment effect'' model of \stcmd{ivqte}.
Computationally, \stcmd{sivqr} implements the smoothed estimator of \citet{KaplanSun2017}, who show that smoothing improves both computation time and statistical accuracy.
Standard errors are computed analytically or by Bayesian bootstrap; for non-i.i.d.\ sampling, \stcmd{sivqr} is compatible with \stcmd{bootstrap}.
I discuss syntax and the underlying methodology, and I compare \stcmd{sivqr} with other commands in an example.

\keywords{\inserttag, sivqr, endogeneity, instrumental variables, quantile regression}
\end{abstract}

\section{Introduction}
\label{sec:intro}

The instrumental variables quantile regression (IVQR) model of \citet{ChernozhukovHansen2005} has become popular for modeling causal effects that differ across individuals (or firms, countries, etc.)\ due to unobserved variables (rather than interactions among observed variables), but the existing IVQR \Stata\ commands \stcmd{ivqreg} \citep{Stata.ivqreg} and \stcmd{ivqreg2} \citep{Stata.ivqreg2} have several limitations.
First, \stcmd{ivqreg} \citep[implementing the estimator of][]{ChernozhukovHansen2006} only allows a single endogenous term.
Besides excluding multiple endogenous regressors, this also excludes interaction or quadratic terms involving the endogenous regressor, for example.
Second, \stcmd{ivqreg2} \citep[implementing the estimator of][]{MachadoSantosSilva2019} imposes a location--scale model that can help if correct but that can result in inconsistency if misspecified.
\Citet[p.\ 159]{MachadoSantosSilva2019} write, ``In these linear models, the validity of [our model] depends on assumptions that are stronger than those required by the IVQR but, when these assumptions are valid, [our estimator] has some potential advantages,'' like ensuring non-crossing of structural quantile functions.
Ideally, \stcmd{sivqr} and \stcmd{ivqreg2} could be combined using a ``model averaging'' approach like in \citet{ChengLiaoShi2019} or \citet{Liu2019}; meanwhile, simply running both may be insightful.
Third, the existing commands' computation time can be long.
%
Fourth, \stcmd{ivqreg} and \stcmd{ivqreg2} share some minor inconveniences, like not supporting syntax for factor variables and interaction terms.
Fifth, \stcmd{ivqreg} is seemingly not actively supported and has additional issues; for example, the newest version 1.0.1 (02dec2010) fixes a problem from version 1.0.0 displaying coefficient labels when using an asterisk in the regressor list (like \stcmd{xvar*}), but it introduces an error when there is overidentification (more excluded instruments than endogenous regressors).

To address these limitations, I introduce the \stcmd{sivqr} command.
The \stcmd{sivqr} command allows models with multiple endogenous terms, supports convenient syntax like for factor variables and interactions terms, and computes a consistent estimator of the IVQR parameters in a reasonable amount of time.
The runtime in the section \ref{sec:ex-main} example is around $25$ seconds for \stcmd{sivqr}, as opposed to a few minutes (\stcmd{ivqreg}) or over $20$ minutes (\stcmd{ivqreg2}).
Although these differences become trivial on small datasets where everything computes quickly, they become even more important when using \stcmd{bootstrap} for cluster-robust standard errors, which requires computing the estimator many times.

The \stcmd{sivqr} command implements the smoothed IVQR estimator of \citet{KaplanSun2017}, using familiar syntax.
Syntax is similar to the \stcmd{ivregress} command (see \rref{ivregress}) but additionally specifying a quantile level.
Likewise, syntax is similar to the \stcmd{qreg} command (see \rref{qreg}) but additionally specifying instruments for the endogenous regressors.
With non-i.i.d.\ sampling, the \stcmd{bootstrap} prefix can be used for standard errors; see \rref{bootstrap} and the example in section \ref{sec:ex}.

Other \Stata\ commands that address endogeneity in quantile models are based on different models than that of \citet{ChernozhukovHansen2005}.
These alternative models have neither ``stronger'' nor ``weaker'' assumptions; they may be more plausible in certain settings and less plausible in others.
The \stcmd{ivqte} command \citep{FrolichMelly2010}
estimates the unconditional (default) or conditional (option \stcmd{aai}) ``local quantile treatment effect,'' analogous to the local average treatment effect model of \citet{ImbensAngrist1994}.
This conditional (\stcmd{aai}) estimand and the IVQR estimand are different but closely related, as explained in section 10.5 of \citet{MellyWuthrich2017} and detailed in \citet{Wuthrich2020}, both of whom compare the models more generally, too.
See also \citet{FrolichMelly2013} for details.
Practically, compared to \stcmd{ivqte} with the \stcmd{aai} option, the main advantage of \stcmd{sivqr} is the ability to handle more than a single, binary endogenous regressor.
The \stcmd{cqiv} command \citep{ChernozhukovEtAl2019,Stata.cqiv} uses a control function estimator based on a triangular model \citep{Lee2007}, also allowing for censoring \citep{ChernozhukovEtAl2015}.
The estimand is the same as for IVQR (see section \ref{sec:interp}).
Practically, compared to \stcmd{cqiv}, the main advantages of \stcmd{sivqr} are handling multiple and/or non-continuous endogenous regressors, as well as allowing simultaneity and reverse causality.
For further comparison of the IVQR and triangular models, see section 9.2.5 of \citet{ChernozhukovHansenWuthrich2017}.

Unfortunately, all these commands (including \stcmd{sivqr}) require ``strong'' instruments for valid standard errors.
In the future, it would be valuable to have \Stata\ implementations of IVQR inference methods robust to weak instruments, such as those of \citet{ChernozhukovHansen2008}, \citet{ChernozhukovEtAl2009}, \citet[\S9.3.3]{ChernozhukovHansenWuthrich2017}, and references therein.
See also \citet{KaplanLiu2021} for a ``$k$-class'' IVQR estimator that can reduce estimation bias from weak instruments, as well as simulation evidence about the usefulness of the conventional ``first-stage'' \fstat\ for assessing instrument strength in IVQR.

Section \ref{sec:gentle} discusses the methodology at a relatively intuitive level.
Section \ref{sec:sivqr} describes syntax and usage of \stcmd{sivqr}.
Section \ref{sec:ex} provides examples that can be replicated with the provided do-file.
Section \ref{sec:methodology} shows some of the theoretical foundations before concluding.
Abbreviations are used for 
instrumental variables (IV), quantile regression (QR), IV quantile regression (IVQR), and two-stage least squares (2SLS).

\section{A gentle introduction to methodology}
\label{sec:gentle}

This section discusses methodology at a relatively non-technical level (compared to section \ref{sec:methodology}).

\subsection{Parameter interpretation}
\label{sec:interp}

First, consider interpretation of the parameters being estimated.
For simplicity, imagine a single regressor $x$ and outcome variable $y$, both scalars.
Scalars $u$ and $v$ represent unobserved variables.

Usually there is a structural model like $y=\beta_0+\beta_1x+v$.
Here, $\beta_0$ and $\beta_1$ are unknown constants and $v$ is everything besides $x$ that causally determines $y$.
Here, $\beta_1$ has a causal interpretation as some effect of $x$ on $y$, but in reality we rarely believe such an effect is the same for all individuals (or firms, or schools, etc.).
There are two approaches: either let differences in effects go into $v$ and interpret $\beta_1$ as some sort of average, or try to learn about the effect heterogeneity.

Consider a structural model that allows individuals to each have their own intercept and slope.
Since the parameters are now individual-specific, they are not constants like $\beta_0$ and $\beta_1$ were, but rather random variables in a ``random coefficients'' model.
The additive error term $v$ from before is simply absorbed into the random intercept, so the structural model is $y=b_0+b_1x$.
That is, each individual has their own $(y,x,b_0,b_1)$, but only $(y,x)$ is observable.

Now imagine the random coefficients can each be written as deterministic functions of a scalar unobservable $u$: $b_0=\beta_0(u)$ and $b_1=\beta_1(u)$.
Because $\beta_0(\cdot)$ and $\beta_1(\cdot)$ are unrestricted, the distribution of $u$ can be normalized to uniform over the unit interval $[0,1]$.
The functions $\beta_0(\cdot)$ and $\beta_1(\cdot)$ are unknown but deterministic; evaluated at a fixed $0<\tau<1$, $\beta_0(\tau)$ and $\beta_1(\tau)$ are unknown constants, just like $\beta_0$ and $\beta_1$ were before.
The differences across individuals are driven by $u$.
Each individual has their own $(y,x,u)$, with $y=\beta_0(u)+\beta_1(u)x$, whereas functions $\beta_0(\cdot)$ and $\beta_1(\cdot)$ are not specific to any individual.

A special case of this random coefficient model is the usual structural model with constant (non-random) coefficients.
Let $\beta_1(u)=\beta_1$, a constant that does not depend on $u$.
Define $v\equiv\beta_0(u)-\beta_0$, so $\beta_0(u)=\beta_0+v$.
The function $\beta_0(\cdot)$ can be interpreted as the inverse CDF (quantile function) of $v$, shifted by $\beta_0$; for example, with $\Phi(\cdot)$ as the standard normal CDF, $v=\Phi^{-1}(u)$ has a standard normal distribution.
Then
\[ y=\beta_0(u)+\beta_1(u)x=\beta_0+v+\beta_1x = \beta_0+\beta_1x+v. \]

Some additional restrictions are required in order to learn about the structural model.
Imagine further that given $x$, $\beta_0(u)+\beta_1(u)x$ is increasing in $u$.
This is known as a ``monotonicity'' assumption.
It also explains why $u$ is often called the ``rank variable'': it describes how somebody's $y$ would rank in the population if everyone were forced to have the same $x$.
For example, somebody with $u=0.5$ would have median $y$, and somebody with $u=0.9$ would have $90$th percentile $y$.
If everybody keeps the same ``rank'' regardless of the $x$ value, then ``rank invariance'' holds.
A weaker assumption called ``rank similarity'' allows the ranking to differ across $x$ as long as the differences are exogenous.

Even if $x$ is endogenous, IVQR can estimate $\beta_0(\tau)$ and $\beta_1(\tau)$ for any $0<\tau<1$ if an instrument $z$ is available that is related to $x$ but independent of $u$ \citep{ChernozhukovHansen2005}.
The interpretation of these parameters depends partly on the rank assumption.
If rank invariance holds, then $\beta_0(\tau)+\beta_1(\tau)x_0$ is the $y$ value that somebody with rank $u=\tau$ would have if we assign them to have value $x=x_0$.
Even with the weaker rank similarity assumption, this is the $\tau$-quantile structural function of \citet[\S3.1]{ImbensNewey2009}: given any $x=x_0$, it provides the $\tau$-quantile of $\beta_0(u)+\beta_1(u)x_0$ over the unconditional population distribution of $u$ (uniform over $[0,1]$), which is $\beta_0(\tau)+\beta_1(\tau)x_0$ due to monotonicity.
Similarly, $\beta_1(\tau)$ can be interpreted as a $\tau$-quantile treatment effect, capturing the difference in the $\tau$-quantile of $y$ between the counterfactual ``untreated'' distribution for which everyone has $x=x_0$ and the counterfactual ``treated'' distribution for which everyone has $x=x_0+1$.

\subsection{Estimation}

\Citet{ChernozhukovHansen2005} show how to derive moment conditions (or ``estimating equations'') to characterize the parameters $\beta_0(\tau)$ and $\beta_1(\tau)$ given a valid instrument and the assumptions discussed above.
For comparison, with $v\equiv y-\beta_0-\beta_1x$ the standard IV moment conditions are
\begin{equation*}
\begin{split}
0&=\E( v)=\E(y-\beta_0-\beta_1x) \\
0&=\E(zv)=\E[z(y-\beta_0-\beta_1x)]
\end{split}
\end{equation*}
The IVQR moment conditions are
\begin{equation}
\label{eqn:IVQR-EE}
\begin{split}
0&=\E[  \Ind{y-\beta_0(\tau)-\beta_1(\tau)x\le0}-\tau]  \\
0&=\E[z(\Ind{y-\beta_0(\tau)-\beta_1(\tau)x\le0}-\tau)]
\end{split}
\end{equation}
where $\Ind{\cdot}$ is the indicator function defined as $\Ind{\mathcal{A}}=1$ if $\mathcal{A}$ is true and otherwise $\Ind{\mathcal{A}}=0$.
This is implied by a conditional quantile restriction on $y-\beta_0(\tau)-\beta_1(\tau)x$ (given $z$), mirroring how the standard IV moments are implied by a conditional mean restriction on $y-\beta_0-\beta_1x$ (given $z$).

Smoothing solves the computational difficulties inherent in \eqref{eqn:IVQR-EE}.
Unlike the standard IV moment conditions, the IVQR moment conditions cannot be solved explicitly for the parameters, nor are the sample moment conditions smooth (differentiable) functions of the parameters.
This computational challenge is addressed by ``smoothing'' the indicator function: replacing it with a function that decreases continuously from $1$ to $0$ rather than discontinuously jumping from $1$ to $0$.
The smoothed function differs from the indicator function only near the jump point; they are identical outside of a window whose width is controlled by a bandwidth (which asymptotically goes to zero).
This smoothing allows the sample system of equations defining the estimator to be solved by standard numerical methods like those available in Mata.
As a bonus, smoothing also improves the statistical properties of the estimator in theory and simulations; see sections 5 and 7 of \citet{KaplanSun2017}.

\section{The sivqr command}
\label{sec:sivqr}

The \stcmd{sivqr} command estimates the coefficients in an instrumental variables quantile regression (IVQR) model, and it provides standard errors.

Syntax, options, and stored results are now shown.
Prefix \stcmd{by} is allowed (see \dref{by}).
After running, postestimation commands like \rref{test} and \rref{predict} can be used as usual.

\subsection{Syntax}

\begin{stsyntax}
sivqr
  \depvar\
  \optional{\varlist$_1$}
  \tt{(}\varlist$_2$ = \varlist$_\textup{iv}$\tt{)}
  \optif\
  \optin\
  \optweight\
  , 
  \dunderbar{q}uantile(\num)
  \optional{\underbar{b}andwidth(\num) \underbar{l}evel(\num) \underbar{r}eps(\num) \underbar{log}iterations \underbar{nocon}stant seed(\num) nodots \underbar{init}ial(\textit{\rmfamily matname})}
\end{stsyntax}

\hangpara
As in \stcmd{ivregress}:
%
\varlist$_1$ has exogenous regressors (or control variables),
%
\varlist$_2$ has endogenous regressors,
%
and \varlist$_\textup{iv}$ has excluded instruments (exogenous variables that instrument for \varlist$_2$).

\hangpara
\stcmd{pweight}s, \stcmd{iweight}s, and \stcmd{fweight}s are allowed; see \uref{11.1.6 weight}.

\subsection{Options}

\hangpara
\texttt{quantile(\num)} specifies the quantile level as in \rref{qreg}: a real number strictly between $0$ and $1$, or alternatively a number between $1$ and $100$ interpreted as a percentile.
For example, \stcmd{quantile(0.5)} specifies the median, as does \stcmd{quantile(50)}.
To prevent inadvertent mistakes, there is no default; this must be specified explicitly.

\hangpara
\texttt{bandwidth(\num)} manually specifies the desired smoothing bandwidth.
Leaving it unspecified invokes a plug-in bandwidth based on \citet{KaplanSun2017} as the default.
If the desired bandwidth is too small to find a numerical solution, then it is increased until a solution is found.
For example, specifying \stcmd{bandwidth(0)} requests the smallest bandwidth that is computationally feasible.
Generally, the default plug-in bandwidth is recommended; compare to an arbitrarily chosen bandwidth, the plug-in usually produces a more accurate estimator (lower mean squared error).
However, for a final analysis, it is prudent to also try a smaller bandwidth like \stcmd{bandwidth(0)} as a sensitivity check (results should be qualitatively similar, though not identical).
A smaller bandwidth tends to yield an estimator with larger variance but smaller (closer to zero) bias.

\hangpara
\texttt{level(\num)} specifies the confidence level, as a percentage, for confidence intervals.
The default is \stcmd{level(95)} or as set by \stcmd{set level}; see \uref{20.7 Specifying the width of confidence intervals}.

\hangpara
\texttt{reps(\num)} specifies the number of bootstrap replications for estimating the variance--covariance matrix and standard errors.
With the default \stcmd{reps(0)}, heteroskedasticity-robust analytic standard errors are reported.
Otherwise, the Bayesian bootstrap of \citet{Rubin1981} is used; it is a valid frequentist bootstrap that also has a nonparametric Bayesian interpretation (see section \ref{sec:methodology-BS}).
%

\hangpara
\texttt{logiterations} prints each iteration of the numerical solver; see \mrefe{solvenl(\,)}, in particular \stcmd{solvenl\_init\_iter\_log(\,)}.
The default is not to print such information, which usually only helps debugging and troubleshooting.

\hangpara
\texttt{noconstant}
omits the intercept term that otherwise is included automatically.

\hangpara
\texttt{seed(\num)}
sets the random-number seed, to make results replicable; default is \stcmd{seed(112358)}.
The current seed is restored at the end of execution.

\hangpara
\texttt{nodots}
suppresses display of the replication dots (see \rref{bootstrap}).

\hangpara
\texttt{initial(\textit{\rmfamily matname})}
sets the initial coefficient values for the numerical search, using the values stored in the matrix (row vector) named \textit{matname}.
If not specified, then \stcmd{qreg} is used to generate initial values.
This option is particularly helpful when estimating the same model across many quantile levels; examples are provided in the \texttt{sivqr\_examples\_init.do} file.

\subsection{Stored results}
\label{sec:sivqr-results}

\stcmd{sivqr} stores the following in \texttt{e()}:
\begin{stresults}
\stresultsgroup{Scalars} \\
\stcmd{e(N)} & number of observations &
\stcmd{e(reps)} & number of bootstrap replications \\
\stcmd{e(bwidth)} & smoothing bandwidth used &
\stcmd{e(q)} & quantile level requested \\
\stcmd{e(bwidth\_req)} & smoothing bandwidth \phantom{requested} requested (or plug-in value) &
\stcmd{e(bwidth\_max)} & maximum plug-in bandwidth \\
\end{stresults}
\begin{stresults}
%
\stresultsgroup{Macros} \\
\stcmd{e(cmd)} & \stcmd{sivqr} &
\stcmd{e(constant)} & \stcmd{noconstant} if specified \\
\stcmd{e(instd)} & instrumented variable(s) &
\stcmd{e(insts)} & instrument(s) \\
\stcmd{e(depvar)} & name of dependent variable &
\stcmd{e(exogr)} & exogenous regressors \\
\stcmd{e(wtype)} & weight type &
\stcmd{e(wexp)} & weight expression \\
\stcmd{e(properties)} & \stcmd{b V} &
\stcmd{e(title)} & title in estimation output \\
\stcmd{e(vcetype)} & \stcmd{Robust} or \stcmd{Bootstrap} &
& \\
\end{stresults}
\begin{stresults}
%
\stresultsgroup{Matrices} \\
\stcmd{e(b)} & estimated coefficient vector &
\stcmd{e(V)} & estimated variance--covariance matrix of the estimator \\
\end{stresults}
\begin{stresults}
%
\stresultsgroup{Functions} \\
\stcmd{e(sample)} & marks estimation sample & & \\
\end{stresults}

Almost all of the above are the same as for the familiar commands \stcmd{ivregress} and/or \stcmd{qreg} (or \stcmd{bsqreg}).
The only exceptions are for the smoothing bandwidth.
By default, the plug-in bandwidth is computed and returned in \stcmd{e(bwidth\_req)}.
More specifically, three plug-in bandwidths are computed (see section \ref{sec:methodology-h}), of which \stcmd{e(bwidth\_req)} is the minimum and \stcmd{e(bwidth\_max)} is the maximum.
If the numerical solver cannot find the solution with \stcmd{e(bwidth\_req)} because it is too close to zero, then the bandwidth is increased until the numerical solver finds the solution.
This feasible bandwidth is returned in \stcmd{e(bwidth)}.
There is nothing wrong with these values being different; for example, specifying \stcmd{bandwidth(0)} makes \stcmd{e(bwidth\_req)} zero and \stcmd{e(bwidth)} the smallest possible bandwidth for which the numerical solver finds a solution.

If the default plug-in bandwidth is used and \stcmd{e(bwidth)} is greater than \stcmd{e(bwidth\_max)}, then there may be deeper problems such as weak instruments.
Although there is currently no weak instrument test specific to IVQR, it would be helpful to run a weak instrument test for 2SLS in this case, like by running \stcmd{ivregress} and then \stcmd{estat firststage}.
If \stcmd{e(bwidth)} is above \stcmd{e(bwidth\_req)} but less than \stcmd{e(bwidth\_max)}, then it may be simply that the plug-in bandwidth was chosen too conservatively (see section \ref{sec:methodology-h}), but it may still be worth running \stcmd{ivregress} and \stcmd{estat firststage}.

\stcmd{sivqr} stores the following in \texttt{r()}:
\begin{stresults}
\stresultsgroup{Scalar} \\
\stcmd{r(level)} & confidence level & & \\
\\\stresultsgroup{Matrix} \\
\stcmd{r(table)} & table of results & & \\
\end{stresults}

These are the standard values stored by \stcmd{ereturn display}; see \pref{ereturn}.

\section{Example: Example 4 from ivregress}
\label{sec:ex}

All results (besides runtimes) can be replicated with the provided \texttt{sivqr\_examples.do} file.
I ran them in \Stata/MP 17.0 with 2 cores \citep{Stata17} on Windows 10, on a few-years-old standard-issue Dell desktop computer (Intel i5-8500 3GHz processor, 8GB RAM).
Besides using version 1.1.0 (22jan2022) of \stcmd{sivqr}, I used version 2.6 (29sep2021) of \stcmd{ivqreg2} from SSC \citep{Stata.ivqreg2}, and version 1.0.0 (28jul2010) of \stcmd{ivqreg} \citep{Stata.ivqreg}.
The most recent version 1.0.1 (02dec2010) from \citet{Stata.ivqregDEC} unfortunately gives an error when trying to use an overidentified model (as in section \ref{sec:ex-main}), ``\texttt{st\_numscalar():  3204  matrix found where scalar required}.''
Other changes in version 1.0.1 are both positive and negative; modifying the section \ref{sec:ex-main} model to only have \stcmd{union} as an excluded instrument, then version 1.0.1 reports standard errors whereas version 1.0.0 does not (instead reporting ``\texttt{Warning:  variance matrix is nonsymmetric or highly singular}''), although the version 1.0.1 runtime is over ten times longer.

The primary focus is comparison with \stcmd{ivqreg} and \stcmd{ivqreg2}, in terms of estimates, speed, and capabilities, including:
\begin{enumerate}
 \item \stcmd{sivqr} with a small bandwidth produces estimates very similar to \stcmd{ivqreg};
 \item \stcmd{ivqreg2} can differ from \stcmd{ivqreg} and \stcmd{sivqr}, though is qualitatively similar;
 \item \stcmd{ivqreg2} can be significantly slower than \stcmd{sivqr}.
\end{enumerate}
As noted earlier, \stcmd{ivqreg} does not allow multiple endogenous regressors, which precludes even a quadratic function of the endogenous regressor, and the factorial operator \stcmd{\#\#} is supported only by \stcmd{sivqr}.

\subsection{Main comparison}
\label{sec:ex-main}

The following revisits Example 4 in \rref{ivregress}.
Briefly, it hopes to estimate a structural wage model, treating job tenure as endogenous; see \rref{ivregress} for details.
I focus on the coefficient on job tenure.

For comparison, \stcmd{ivregress 2sls} is run first.
This provides a helpful ``sanity check'': even if there are differences across quantile levels, the IVQR coefficient estimates and signs should still be similar overall, especially around the median.

\begin{stlog}
. webuse nlswork , clear
(National Longitudinal Survey of Young Women, 14-24 years old in 1968)
{\smallskip}
. ivregress 2sls ln_wage c.age\#\#c.age birth_yr grade (tenure = union wks_work msp),
>   vce(cluster idcode)
{\smallskip}
Instrumental variables 2SLS regression            Number of obs   =     18,625
                                                  Wald chi2(5)    =    1600.81
                                                  Prob > chi2     =     0.0000
                                                  R-squared       =          .
                                                  Root MSE        =     .48441
{\smallskip}
                             (Std. err. adjusted for 4,110 clusters in idcode)
\HLI{13}{\TOPT}\HLI{64}
             {\VBAR}               Robust
     ln_wage {\VBAR} Coefficient  std. err.      z    P>|z|     [95\% conf. interval]
\HLI{13}{\PLUS}\HLI{64}
      tenure {\VBAR}   .1060832   .0044835    23.66   0.000     .0972957    .1148707
         age {\VBAR}   .0162345   .0069467     2.34   0.019     .0026193    .0298497
             {\VBAR}
 c.age\#c.age {\VBAR}  -.0005309   .0001155    -4.59   0.000    -.0007573   -.0003044
             {\VBAR}
    birth_yr {\VBAR}  -.0091139   .0022602    -4.03   0.000    -.0135438   -.0046839
       grade {\VBAR}    .070454   .0031158    22.61   0.000     .0643471    .0765609
       _cons {\VBAR}   .9079537   .1681148     5.40   0.000     .5784548    1.237453
\HLI{13}{\BOTT}\HLI{64}
Instrumented: tenure
 Instruments: age c.age\#c.age birth_yr grade union wks_work msp
\end{stlog}

The following shows results from \stcmd{sivqr} at the median level.
To cluster standard errors at the individual level, the \stcmd{bootstrap} prefix is used after using \stcmd{xtset} appropriately.%
\footnote{For example, see \url{https://www.stata.com/support/faqs/statistics/bootstrap-with-panel-data/}}
With \stcmd{bootstrap}, it would only waste time (without changing the reported estimates or standard errors) to specify the \stcmd{reps} option (with a non-zero number) for \stcmd{sivqr}.
At the median, the \stcmd{sivqr} coefficient estimates (other than the intercept) as well as the corresponding cluster-robust standard errors are similar to those from \stcmd{ivregress 2sls}.

\begin{stlog}
. generate newid = idcode
{\smallskip}
. xtset newid year
{\smallskip}
Panel variable: newid (unbalanced)
 Time variable: year, 68 to 88, but with gaps
         Delta: 1 unit
{\smallskip}
. bootstrap , reps(100) cluster(idcode) idcluster(newid) seed(112358) : sivqr ln_
> wage c.age\#\#c.age birth_yr grade (tenure = union wks_work msp) , quantile(0.50)
(running {\bftt{sivqr}} on estimation sample)
{\smallskip}
Bootstrap replications (100)
\HLI{4}{\PLUS}\HLI{3} 1 \HLI{3}{\PLUS}\HLI{3} 2 \HLI{3}{\PLUS}\HLI{3} 3 \HLI{3}{\PLUS}\HLI{3} 4 \HLI{3}{\PLUS}\HLI{3} 5 
..................................................    50
..................................................   100
{\smallskip}
Smoothed instrumental variables quantile regression (SIVQR)   Quantile = .5
Smoothing bandwidth used = .0600669                 Number of obs =     18,625
                              (Replications based on 4,110 clusters in idcode)
\HLI{13}{\TOPT}\HLI{64}
             {\VBAR}   Observed   Bootstrap                         Normal-based
     ln_wage {\VBAR} coefficient  std. err.      z    P>|z|     [95\% conf. interval]
\HLI{13}{\PLUS}\HLI{64}
      tenure {\VBAR}   .1076941   .0046079    23.37   0.000     .0986627    .1167255
         age {\VBAR}   .0060803   .0073372     0.83   0.407    -.0083005     .020461
             {\VBAR}
 c.age\#c.age {\VBAR}  -.0003585   .0001212    -2.96   0.003     -.000596   -.0001209
             {\VBAR}
    birth_yr {\VBAR}   -.011967   .0021773    -5.50   0.000    -.0162344   -.0076997
       grade {\VBAR}    .065723   .0030378    21.63   0.000      .059769     .071677
       _cons {\VBAR}   1.255391   .1643564     7.64   0.000     .9332579    1.577523
\HLI{13}{\BOTT}\HLI{64}
Instrumented:  tenure
Instruments:   age c.age\#c.age birth_yr grade union wks_work msp
\end{stlog}

The following code produces the results in tables \ref{tab:ex-ivregress4-est} and \ref{tab:ex-ivregress4-runtime}, besides the cluster bootstrap standard errors computed above (that are repeated for other quantiles).
Compared to the provided do-file, below I omit various \stcmd{timer} and matrix storage commands (and output).
I run \stcmd{sivqr} with both the default plug-in bandwidth as well as \stcmd{bandwidth(0)}, which tries to find the smallest computationally feasible bandwidth.
The squared age variable is generated manually because \stcmd{ivqreg} and \stcmd{ivqreg2} do not support the factorial operator \stcmd{\#\#}.

\begin{stlog}
. generate agesq = age{\caret}2
(24 missing values generated)
{\smallskip}
. forvalues i = 1/3 {\lbr}
  2.   local q = `i'/4
  3.   sivqr ln_wage age agesq birth_yr grade (tenure = union wks_work msp) ,
>   quantile(`q')
  4.   sivqr ln_wage age agesq birth_yr grade (tenure = union wks_work msp) ,
>   quantile(`q') bandwidth(0)
  5.   ivqreg ln_wage age agesq birth_yr grade (tenure = union wks_work msp) ,
>   q(`q') robust
  6. {\rbr}
{\smallskip}
. ivqreg2 ln_wage tenure age agesq birth_yr grade , quantile(0.25 0.5 0.75) 
>   instruments(union wks_work msp age agesq birth_yr grade)
\end{stlog}

\sisetup{detect-mode,
        tight-spacing           = true,
        group-digits            = false,
        input-signs             = ,
        input-symbols           = ,
        input-open-uncertainty  = ,
        input-close-uncertainty = ,
        table-align-text-pre    = false,
        round-mode              = places,
        round-precision         = 4,
        table-space-text-pre    = (,
        table-space-text-post   = ),
  table-number-alignment=center, table-format=1.4 }
\begin{table}[htbp]
\caption{Estimates of \stcmd{tenure} coefficient; clustered standard errors.}
\label{tab:ex-ivregress4-est}
\begin{center}
\begin{tabular}{cSSSS}
\hline
\noalign{\smallskip}
Quantile 
&  \multicolumn{1}{c}{\stcmd{sivqr}}
&  \multicolumn{1}{c}{\stcmd{sivqr, b(0)}}
& \multicolumn{1}{c}{\stcmd{ivqreg}}
& \multicolumn{1}{c}{\stcmd{ivqreg2}} \\ 
\noalign{\smallskip}
\hline
\noalign{\smallskip}
$0.25$ 
& 0.0865756 & 0.0860257 
& 0.0861186 & 0.0933732 \\
& (0.0031621)
&&& \\
$0.50$
& 0.1076941 & 0.1080343
& 0.1032661 & 0.1084423 \\
& (0.0046079)
&&& \\
$0.75$
& 0.1565857 & 0.1553029
& 0.1584188 & 0.1250514 \\
& (0.0103318)
&&& \\
\noalign{\smallskip}
\hline
\noalign{\smallskip}
\multicolumn{5}{l}{\textbf{Note:} 2SLS estimate (std.\ err.)\ is $0.1061$ ($0.0045$).}
\end{tabular}
\end{center}
\end{table}

Table \ref{tab:ex-ivregress4-est} shows the different \stcmd{tenure} coefficient estimates at different quantiles, with the 2SLS results for comparison.
To help gauge the magnitude of the differences, table \ref{tab:ex-ivregress4-est} also shows cluster-robust standard errors from 2SLS and from using \stcmd{bootstrap} with \stcmd{sivqr}.
(None of the commands produces cluster-robust standard errors by itself, so the default standard errors are not reported; they are generally too small here, as expected.)

At the median, all estimates are similar.
All are well within one standard error of the 2SLS estimate.
(The 2SLS estimate can differ greatly from median IVQR estimates, just as the sample mean and median can differ greatly, but they are often similar.)
The cluster-bootstrap \stcmd{sivqr} standard error is also similar to the 2SLS cluster-robust standard error.

Away from the median, \stcmd{sivqr} (either bandwidth) and \stcmd{ivqreg} remain similar to each other, but other differences arise.
First, the IVQR estimates are all increasing with the quantile level, differing significantly from the 2SLS estimate both economically and statistically.
Compared to the $0.25$ quantile estimates, most of the $0.75$ quantile estimates are nearly twice as large, suggesting substantial heterogeneity.
This could be interpreted as individuals with higher levels of ``ability'' (and other unobserved wage determinants) having higher returns to job tenure.

Second, \stcmd{ivqreg2} differs from the other IVQR estimators.
This is especially seen at the $0.75$ quantile: the \stcmd{ivqreg2} estimate of the return to tenure is below the other three estimates by around $0.03$ (three percentage points), which is around a three standard error difference.
It is possible that the location--scale model of \stcmd{ivqreg2} is correctly specified and that this difference is only due to larger estimation error of \stcmd{sivqr} and \stcmd{ivqreg}, but given the consistency of both \stcmd{sivqr} and \stcmd{ivqreg} along with the difference being three standard errors, it seems likely that at least part of the difference is explained by misspecification of the \stcmd{ivqreg2} location--scale model.

As a side note, these \stcmd{ivqreg2} results were using the regressor order that gave the most economically plausible estimates.
Keeping everything the same but with \stcmd{tenure} last in the regressor list, the estimates of the \stcmd{tenure} coefficient become significantly negative, and two warning messages appear, ``\texttt{WARNING: .03758389\% of the fitted values of the scale function are not positive}'' and ``\texttt{Warning:  variance matrix is nonsymmetric or highly singular}'' (with missing standard errors).

\begin{table}[htbp]
\caption{Runtime (seconds) of IVQR commands on \stcmd{nlswork} data.}
\label{tab:ex-ivregress4-runtime}
\begin{center}
\begin{tabular}{crrrrr}
\hline
\noalign{\smallskip}
Quantile
& \stcmd{bootstrap:sivqr} & \stcmd{sivqr} & \stcmd{sivqr, b(0)} & \stcmd{ivqreg} & \stcmd{ivqreg2} \\ 
\noalign{\smallskip}
\hline
\noalign{\smallskip}
$0.25$ &  $861$ & $16$ &  $98$ & $130$ & $\smash{\uparrow}\phantom{64}$ \\
$0.50$ &  $500$ &  $5$ &  $68$ &  $68$ & $1304$ \\
$0.75$ & $1026$ &  $4$ &  $47$ & $153$ & $\smash{\downarrow}\phantom{64}$ \\
\noalign{\smallskip}
\hline
\end{tabular}
\end{center}
\end{table}

Table \ref{tab:ex-ivregress4-runtime} shows the runtime of the different commands at different quantiles.
The fastest is \stcmd{sivqr} with the default plug-in bandwidth and analytic standard errors, taking around $25$ seconds to compute at all three quantiles.
Using \stcmd{bandwidth(0)} is significantly slower (a few minutes total, instead of seconds) because it requires many more calls to the numerical solver, trying to find the very smallest usable bandwidth.
The \stcmd{ivqreg} time is yet slower than \stcmd{bandwidth(0)}, also multiple minutes total.
Naturally, running a cluster bootstrap with $100$ replications takes even longer, even using the fast \stcmd{sivqr} with plug-in bandwidth, taking around $40$ minutes total.
Even without bootstrap, \stcmd{ivqreg2} takes over $20$ minutes, an order of magnitude more than the $25$ seconds taken by \stcmd{sivqr}.

\subsection{More computation times}
\label{sec:ex-more}

The provided do-file illustrates some additional important considerations about computation time.
Most of the \stcmd{sivqr} computation time is from calling the \stcmd{solvenl} numerical solver.
Naturally, some numerical problems are more difficult to solve quickly.
In particular, difficulty is higher with less smoothing or when there is less data near the quantile of interest (roughly speaking), which most often happens farther into the tails at quantile levels near zero or one.

Table \ref{tab:ex-ivregress4-time-q} illustrates this by running \stcmd{sivqr} over quantiles $0.10,0.15,\ldots,0.90$, using either the plug-in bandwidth (always between $0.05$ and $0.08$) or \stcmd{bandwidth(0)} (always between $0.00001$ and $0.00012$, except at the $0.9$ quantile).
First, although there are other factors, runtime is generally longer in the tails: with the plug-in bandwidth, runtime is at least $8$ seconds at quantile levels $\{0.10,0.80,0.85,0.90\}$ but under $8$ seconds in between.
The pattern is less clear for \stcmd{bandwidth(0)} because it depends not only on the runtime per \stcmd{solvenl} call, but also on the number of such calls.
Second, the runtime is longer with the smaller bandwidth at $16$ out of $17$ quantile levels.

\begin{table}[htbp]
\caption{Runtime (seconds) across quantile levels, \stcmd{nlswork} data.}
\label{tab:ex-ivregress4-time-q}
\begin{center}
\begin{tabular}{crr}
\hline
\noalign{\smallskip}
Quantile & plug-in & \stcmd{bandwidth(0)} \\ 
\noalign{\smallskip}
\hline
\noalign{\smallskip}
$0.10$ & $16$ &  $99$ \\
$0.15$ &  $3$ &  $51$ \\
$0.20$ &  $6$ &  $49$ \\
$0.25$ &  $5$ &  $88$ \\
$0.30$ &  $2$ &  $82$ \\
$0.35$ &  $2$ &  $54$ \\
$0.40$ &  $3$ &  $46$ \\
$0.45$ &  $4$ &  $25$ \\
$0.50$ &  $4$ &  $61$ \\
$0.55$ &  $4$ &  $29$ \\
$0.60$ &  $3$ &  $27$ \\
$0.65$ &  $2$ &  $29$ \\
$0.70$ &  $5$ &  $38$ \\
$0.75$ &  $4$ &  $44$ \\
$0.80$ & $13$ &  $36$ \\
$0.85$ & $53$ &  $44$ \\
$0.90$ &  $8$ &  $43$ \\
\noalign{\smallskip}
\hline
\end{tabular}
\end{center}
\end{table}

Also, even with the same original dataset, some bootstrap datasets are naturally more difficult than others, so bootstrap runtime can depend on the random seed.
With \stcmd{quantile(0.5)}, out of $100$ runs with different seeds, runtimes for the built-in \stcmd{sivqr} bootstrap with $20$ replications ranged from $9$ seconds to $30$ seconds, with lower and upper quartiles $11$ seconds and $16$ seconds.
Among five runs with the \stcmd{bootstrap} prefix (also $20$ replications), runtimes ranged from $79$ to $90$ seconds.
(I conjecture this is much slower because it resamples a new dataset rather than merely reweighting the original dataset like the Bayesian bootstrap of \stcmd{sivqr}, but I have not broken down the runtime by function to investigate further.)
However, it cannot be known beforehand which seeds are ``faster'' for a particular dataset, so using the default seed is recommended, which also avoids any appearance of trying to manipulate results.

\section{Methods and formulas}
\label{sec:methodology}

This section contains additional theoretical details.
After introducing notation in section \ref{sec:methodology-notation}, section \ref{sec:methodology-est} concerns the difficulty of estimation and the solution of \citet{KaplanSun2017}, while section \ref{sec:methodology-ID} has some of the identification arguments from \citet{ChernozhukovHansen2005} that characterize causal parameters as solutions to moment conditions (estimating equations).
Section \ref{sec:methodology-h} describes various bandwidths.
Section \ref{sec:methodology-BS} has brief notes on the Bayesian bootstrap used to compute standard errors.

These details may provide a deeper understanding for some readers, but they may also be skipped without hindering successful application of \stcmd{sivqr} in practice.

\subsection{Notation}
\label{sec:methodology-notation}

Notationally, let $y$ be the scalar outcome variable, $\vecf{x}$ a column vector of regressors, and $u$ a scalar unobserved variable.
Let $\vecf{d}$ be a subset of $\vecf{x}$ containing all the endogenous regressors.
Let $\vecf{z}$ be a column vector of all exogenous variables: both exogenous regressors in $\vecf{x}$ as well as the excluded instruments (instrumental variables).
The scalar random variable $u$ is unobserved.
Following \textit{\sj} convention, vectors are lowercase bold (like $\vecf{x}$) and scalars are lowercase plain (like $y$) since uppercase is reserved for matrices, but with some luck, random variables can be distinguished from their possible non-random realizations.

Additional notation is introduced as needed below.

\subsection{Estimation}
\label{sec:methodology-est}

First, as a helpful reference point for intuition, recall the standard (``mean'') instrumental variables (IV) model and estimator.
The goal is to estimate the non-random parameter vector $\vecf{\beta}$ in the structural model
\begin{equation*}
y = \vecf{x}'\vecf{\beta} + u
\end{equation*}
If the instruments $\vecf{z}$ are independent of the structural error $u$, then
\begin{equation}
\label{eqn:EZU0}
\vecf{0} = \E(u\mid\vecf{z})
\implies
\vecf{0} = \E(\vecf{z}u)
= \E[\vecf{z}(y-\vecf{x}'\vecf{\beta})] 
\end{equation}
where the $\implies$ follows from the law of iterated expectations.
If the dimensions of $\vecf{z}$ and $\vecf{x}$ are the same (``exact identification''), then this can be solved explicitly for $\vecf{\beta}$:
\begin{equation*}
\E(\vecf{z}y) = \E(\vecf{z}\vecf{x}'\vecf{\beta})
\implies
\E(\vecf{z}y) = \E(\vecf{z}\vecf{x}') \vecf{\beta}
\implies
\vecf{\beta} = [\E(\vecf{z}\vecf{x}')]^{-1} \E(\vecf{z}y) 
\end{equation*}
and (with i.i.d.\ data) the expectations are replaced by sample averages to yield the familiar IV estimator.

Superficially, imagine replacing the conditional mean restriction $\E(y\mid\vecf{z})$ in \eqref{eqn:EZU0} with a conditional $\tau$-quantile restriction for some $0<\tau<1$:
\begin{equation}
\label{eqn:QZU0}
0 = \Q_\tau(u\mid \vecf{z}) 
\end{equation}
That is, conditional on any value of $\vecf{z}$, the conditional distribution of $u$ has its $\tau$-quantile equal to zero.
By definition, the $\tau$-quantile of a distribution is the value with $\tau$ probability below that value, and the same is true conditional on $\vecf{z}$.
Thus, \eqref{eqn:QZU0} is equivalent to
\begin{equation*}
\tau = \Pr(u\le0\mid\vecf{z})
= \Pr(y-\vecf{x}'\vecf{\beta}\le0\mid\vecf{z}) 
\end{equation*}
Let $\Ind{\cdot}$ be the indicator function such that $\Ind{\mathcal{A}}=1$ if $\mathcal{A}$ is true and otherwise $\Ind{\mathcal{A}}=0$.
Rewriting $\Pr(\cdot)$ as $\E[\Ind{\cdot}]$ and applying the law of iterated expectations as in \eqref{eqn:EZU0},
\begin{align}\notag
\tau = \E[\Ind{y-\vecf{x}'\vecf{\beta}\le0}\mid\vecf{z}]
&\implies
0 = \E[\Ind{y-\vecf{x}'\vecf{\beta}\le0}-\tau\mid\vecf{z}]
\\
&\implies
\vecf{0} = \E[\vecf{z}(\Ind{y-\vecf{x}'\vecf{\beta}\le0} - \tau)] 
\label{eqn:EZ10}
\end{align}

Despite some similarities, computing an IVQR estimator based on \eqref{eqn:EZ10} is much more difficult than computing the mean IV estimator based on \eqref{eqn:EZU0}.
With mean IV, it was possible to solve for $\vecf{\beta}$ and replace the mean (expectation) with the sample average.
Here, the $\vecf{\beta}$ is stuck inside the indicator function $\Ind{\cdot}$, so it cannot be solved for explicitly.
Further, it may be impossible to solve the equation exactly after substituting in the sample average.

These difficulties are both addressed by ``smoothing'' the indicator function in \eqref{eqn:EZ10}.
Replacing $\Ind{\cdot}$ with a very similar but continuously differentiable function allows the sample system of equations to be solved quickly by standard numerical methods.
Specifically, replacing $\Ind{v\le0}$, $\tilde{I}(v)$ is a function of $v$ that smoothly decreases from $1$ to $0$ over $-1\le v\le1$ instead of decreasing discontinuously from $1$ to $0$ at $v=0$.%
\footnote{This notation is simpler and equivalent to (but different than) that in \citet{KaplanSun2017}.}
Adding a bandwidth $h$ allows $\tilde{I}(v/h)$ to decrease more steeply over $-h\le v\le h$.
The estimator $\sthat{\vecf{\beta}}$ solves the ``smoothed estimating equations'' (or ``smoothed moment conditions'')
\begin{equation}
\label{eqn:SEE}
\vecf{0} = \frac{1}{n}\sum_{i=1}^{n}
  \vecf{z}_i[\tilde{I}((y_i-\vecf{x}_i'\sthat{\vecf{\beta}})/h) - \tau] 
\end{equation}
As a bonus, this smoothing also improves estimation precision, as shown both theoretically and in simulations by \citet[\S\S5,7]{KaplanSun2017}.
Weights $w_i$ can be inserted into \eqref{eqn:SEE} as needed:
\begin{equation}
\label{eqn:SEEw}
\vecf{0} = \frac{1}{n}\sum_{i=1}^{n}
  w_i \vecf{z}_i[\tilde{I}((y_i-\vecf{x}_i'\sthat{\vecf{\beta}})/h) - \tau] 
\end{equation}

Specifically, \stcmd{sivqr} uses a piecewise linear $\tilde{I}(\cdot)$ that connects the smoothed IVQR estimator to other estimators.
Instead of jumping down from $1$ to $0$ discontinuously, $\tilde{I}(\cdot)$ has $\tilde{I}(v)=1$ for $v\le-1$ and $\tilde{I}(v)=0$ for $v\ge1$, but transitions linearly with $\tilde{I}(v)=(1-v)/2$ for $-1<v<1$.
\Citet[\S2.2, p.\ 111]{KaplanSun2017} show how this choice produces the Winsorized mean estimator of the type in \citet[ex.\ (iii), p.\ 79]{Huber1964} in the special case with $\tau=0.5$ and an intercept-only model ($\vecf{x}=\vecf{z}=1$).
They also show (in Section 2.2) how using a very large amount of smoothing ($h$) turns the smoothed IVQR estimator into the usual (mean) IV estimator, with an adjusted intercept.
Intuitively, if $h$ is large enough that every observation is smoothed, then $\tilde{I}(\cdot)$ is a linear function of the residuals $y_i-\vecf{x}_i'\sthat{\vecf{\beta}}$, just as in the mean IV moment conditions; if $\tau\ne0.5$, then the intercept is different, but the slope estimates are identical to the IV slope estimates for any $\tau$.
Of course, in practice $h$ is small, but this shows that the worst-case effect of choosing $h$ too large is that you simply get the usual IV slope estimates.

If there are more instruments than parameters (overidentification), then a different $\vecf{z}$ is used in \eqref{eqn:SEE}.
Specifically, it is replaced by the linear projection of the regressor vector $\vecf{x}$ onto the instruments $\vecf{z}$, which has the same dimension as $\vecf{x}$ and thus $\vecf{\beta}$.
This linear projection is motivated by the two-stage least squares estimator, which can also be written as the mean IV estimator when replacing $\vecf{z}$ with the linear projection of $\vecf{x}$ onto $\vecf{z}$.
Although theoretically more efficient estimators may exist, this estimator remains consistent, reliable, and fast to compute regardless of the degree of overidentification.

\subsection{Identification}
\label{sec:methodology-ID}

It remains to motivate the conditional quantile restriction in \eqref{eqn:QZU0} from a causal model, as originally done by \citet{ChernozhukovHansen2005}.%
\footnote{See also Chapter 6 of \citet{Kaplan2020distnp} for an introductory discussion.}
Instead of assuming every individual (or firm, or school, etc.)\ has the same coefficients $\vecf{\beta}$, imagine each individual has their own coefficient vector $\vecf{b}$.
This is a ``random coefficient'' model, meaning $\vecf{b}$ can differ by individual the same way that $y$, $\vecf{x}$, and $\vecf{z}$ do:
\begin{equation*}
y = \vecf{x}'\vecf{b} 
\end{equation*}
An additional error term would be redundant; for example, if $y=b_0+b_1x+v$, then the random intercept can simply absorb $v$ to become $b_0+v$.
To make the model more tractable empirically, imagine $\vecf{b}$ can be written as a deterministic (but unknown) vector-valued function $\vecf{\beta}(\cdot)$ applied to a scalar unobserved $u$:
\begin{equation}
\label{eqn:IVQR-structural}
y = \vecf{x}'\vecf{\beta}(u) 
\end{equation}
Assuming $u$ is continuous, it can be normalized to have a $\UnifDist(0,1)$ distribution (uniformly distributed between $0$ and $1$): any transformation is simply absorbed into $\vecf{\beta}(\cdot)$.
Assume a ``monotonicity'' property that given any value of $\vecf{x}$, $\vecf{x}'\vecf{\beta}(u)$ is an increasing function of $u$, and like before assume the instruments are independent of $u$.
Then,
\begin{align*}
\Pr(y\le\vecf{x}'\vecf{\beta}(\tau)\mid\vecf{z})
&=
\Pr(\vecf{x}'\vecf{\beta}(u)\le\vecf{x}'\vecf{\beta}(\tau)\mid\vecf{z})
&&\textrm{by \eqref{eqn:IVQR-structural}}
\\&=
\Pr(u\le\tau\mid\vecf{z})
&&\textrm{by monotonicity}
\\&=
\Pr(u\le\tau)
&&\textrm{by }u\independent\vecf{z}
\\&=
\tau
&&\textrm{by }u\sim\UnifDist(0,1)
\end{align*}
That is, for any $u=\tau$, $\vecf{\beta}(\tau)$ is identified by the conditional quantile restriction from \eqref{eqn:QZU0}, which can be estimated by the \stcmd{sivqr} command using \eqref{eqn:SEE}.
Given $\tau$, the function $q_\tau(\vecf{x})=\vecf{x}'\vecf{\beta}(\tau)$ is also the $\tau$-quantile structural function introduced by \citet[\S3.1]{ImbensNewey2009}.

\subsection{Plug-in bandwidth}
\label{sec:methodology-h}

This section only applies to the plug-in bandwidth.
If instead the \stcmd{bandwidth} option is manually specified, then none of the following applies.

\Citet[prop.\ 2, p.\ 117]{KaplanSun2017} provide a theoretical ``optimal'' smoothing bandwidth.
Specifically, it minimizes the asymptotic mean squared error of the smoothed estimating equations (moment conditions) themselves.
It also minimizes the asymptotic mean squared error of a particular linear combination of the estimated coefficients, although it does not do so for every possible linear combination; see section 5 of \citet{KaplanSun2017}.

A plug-in (data-dependent) version of the optimal bandwidth from \citet{KaplanSun2017} is implemented in \stcmd{sivqr} as follows.
First, as in their proposition 2, the bandwidth is simplified by assuming $v\equiv y-\vecf{x}'\vecf{\beta}(\tau)$ is independent of the full vector of instruments.
Second, the smoothed indicator function is piecewise linear, as mentioned in section \ref{sec:methodology-est}.
In the notation of \citet{KaplanSun2017},
\begin{equation}
\label{eqn:G-linear}
G(v) = \max\{0, \min\{1, (v+1)/2 \} \} 
\end{equation}
%
Third, the smallest amount of underlying smoothness in the data-generating process is assumed, $r=2$ \citep[see][ass.\ 3]{KaplanSun2017}.

With the chosen $G(\cdot)$ in \eqref{eqn:G-linear} and $r=2$, simplifying the optimal bandwidth $h^*$ from propositon 2 of \citet{KaplanSun2017} with $v\independent\vecf{z}$,
\begin{gather*}
1-\int_{-1}^{1}[G(v)]^2\diff{v}
= 1-\int_{-1}^{1}(v+1)^2/4\diff{v}
 = 1/3  \\
\left[\int_{-1}^{1}G'(v)v^r\diff{v}\right]^2
= \left[\int_{-1}^{1}v^2/2\diff{v}\right]^2
= 1/9  \\
\begin{split}
h^* &= 
  \left(\frac{(r!)^2\left[1-\int_{-1}^{1}[G(v)]^2\diff{v}\right]f_v(0)}{2r\left[\int_{-1}^{1}G'(v)v^r\diff{v}f_v^{(r-1)}(0)\right]^2} \frac{d}{n}\right)^{1/(2r-1)}
\\&= 
 n^{-1/3}
  \left(d\frac{4(1/3)f_v(0)}{4(1/9)\left[f_v'(0)\right]^2}\right)^{1/3}
=  n^{-1/3}
  \left(3df_v(0) / [f_v'(0)]^2\right)^{1/3}
\end{split}
\end{gather*}
where $f_v(\cdot)$ is the probability density function of $v$ and $f_v''(\cdot)$ its second derivative, and $d$ is the dimension (length) of $\vecf{\beta}(\tau)$.

The \stcmd{sivqr} plug-in bandwidth is the minimum of a few different values.
By taking the minimum (instead of average or maximum), mistakes tend to be in the direction of undersmoothing compared to the ``optimal'' smoothing that minimizes mean squared error.
This results in lower bias (but higher variance) than is optimal, because smoothing reduces variance at the cost of increased bias.
The minimum is taken among a nonparametrically estimated version of $h^*$, a Gaussian reference version, and the Silverman rule of thumb bandwidth \citep[\S3.4.2]{Silverman1986} for estimating the density of $v$, as suggested by \citet{FernandesGuerreHorta2021} for non-IV smoothed QR.
The former two are detailed further below.

After finally using the plug-in bandwidth to compute the smoothed IVQR estimator, the residuals $\sthat{v}_i$ are then recomputed from the new $\sthat{\vecf{\beta}}(\tau)$, leading to an updated plug-in bandwidth, and the resulting IVQR estimate is reported by \stcmd{sivqr}.

\subsubsection{Nonparametrically estimated bandwidth}

One approach is to replace $f_v(0)$ and $f_v'(0)$ in $h^*$ by nonparametric kernel estimates.
Since $v$ is unobserved, an initial estimate $\sthat{\vecf{\beta}}(\tau)$ must be used to construct residuals
\begin{equation*}
\sthat{v}_i = y_i - \vecf{x}_i'\sthat{\vecf{\beta}}(\tau) 
\end{equation*}

For $f(0)$ (dropping the $v$ subscript for now), given the $\sthat{v}_i$, the usual kernel density estimator is
\begin{equation}
\label{eqn:f0-hat}
\frac{1}{ns}\sum_{i=1}^{n}K(-\sthat{v}_i/s)
\end{equation}
where $s$ is the kernel bandwidth (not to be confused with the IVQR smoothing bandwidth $h$) and the kernel function $K(\cdot)$ is chosen to be the Gaussian kernel.
For $s$, a pointwise version of Silverman's rule of thumb \citep[\S3.4.2]{Silverman1986} is used because interest is in $f(0)$, not the full function $f(\cdot)$, so it is better to minimize the pointwise mean squared error than the integrated mean squared error.
The (asymptotic) pointwise mean squared error and optimal bandwidth can be found in \citet[Ch.\ 32, p.\ 536]{DasGupta2008}, for example:
\begin{equation*}
s^* = n^{-1/5}\left(\frac{f(0)}{\{f''(0)\}^2}\right)^{1/5}
      \left( \frac{\int_\R\{K(v)\}^2\diff{v}}{\{\int_\R K(v)v^2\diff{v}\}^2}\right)^{1/5}
\end{equation*}

Following the Gaussian reference approach of \citet[\S3.4.2]{Silverman1986}, assume $f(\cdot)$ is the density of a $\NormDist(\mu,\sigma^2)$ distribution; that is, $f(x)=\phi((x-\mu)/\sigma)/\sigma$, and the corresponding cumulative distribution function is $F(x)=\Phi((x-\mu)/\sigma)$, where $\phi(\cdot)$ and $\Phi(\cdot)$ are respectively the density and distribution functions of the standard normal distribution.
Given the restriction that the $\tau$-quantile of $v$ is zero,
\begin{equation*}
\tau = \Phi((0-\mu)/\sigma)
\implies -\mu/\sigma = \Phi^{-1}(\tau)
\end{equation*}
Thus, the density at zero is
\begin{equation}
\label{eqn:f0}
f(0)=\phi((0-\mu)/\sigma)/\sigma
    =\phi(\Phi^{-1}(\tau))/\sigma 
\end{equation}

Similarly, assuming normality allows us to express $\sthat{f}''(0)$ in terms of $\sthat\sigma$.
The second derivative of the $\NormDist(\mu,\sigma^2)$ density evaluated at zero is%
\footnote{\texttt{D[PDF[NormalDistribution[\textbackslash[Mu], \textbackslash[Sigma]], x], \{x,2\}]} at \url{http://www.wolframalpha.com} for example; then plug in $x=0$ and $-\mu/\sigma=\Phi^{-1}(\tau)$.}
\begin{align*}
f''(0) &= f(0)\{(0-\mu)^2/\sigma^4-1/\sigma^2\}
= \overbrace{\phi(\Phi^{-1}(\tau))\sigma^{-1}}^{=f(0)\textrm{ by \eqref{eqn:f0}}}
  \sigma^{-2}[\{\Phi^{-1}(\tau)\}^2-1]
\\&= \sigma^{-3}
  \phi(\Phi^{-1}(\tau))
  [\{\Phi^{-1}(\tau)\}^2-1]
\end{align*}
Thus,
\begin{align*}
\frac{f(0)}{\{f''(0)\}^2}
= \frac{\phi(\Phi^{-1}(\tau)) \sigma^{-1}}%
       {\sigma^{-6} \{\phi(\Phi^{-1}(\tau))\}^2
        [\{\Phi^{-1}(\tau)\}^2-1]^2}
= \frac{\sigma^5}{\{\phi(\Phi^{-1}(\tau))\}
        [\{\Phi^{-1}(\tau)\}^2-1]^2}
\end{align*}
Also, for the Gaussian kernel,\footnote{Unfortunately, the value in table 32.1 in \citet[p.\ 537]{DasGupta2008} is incorrectly stated as $1/\sqrt{2\pi}$.}
\begin{align}
\notag
\int_\R\{K(v)\}^2\diff{v}
&= \frac{1}{2\pi}\int_\R\exp(-v^2)\diff{v}
= 1/(2\sqrt{\pi}) , \\
\int_\R K(v)v^2\diff{v} &= 1
\label{eqn:mu2}
\end{align}
Finally, plugging everything into the formula for $s^*$,
\begin{align*}
s^* &= n^{-1/5}
      \left(\frac{f(0)}{\{f''(0)\}^2}\right)^{1/5}
      \left( \frac{\int_\R\{K(v)\}^2\diff{v}}{\{\int_\R K(v)v^2\diff{v}\}^2}\right)^{1/5}
\\&= n^{-1/5} \sigma
     \left[ \phi(\Phi^{-1}(\tau))
        [\{\Phi^{-1}(\tau)\}^2-1]^2 \right]^{-1/5}
     \left(\frac{1/(2\sqrt{\pi})}{1^2}\right)^{1/5}
\\&= 0.776 n^{-1/5} \sigma
     \left[ \phi(\Phi^{-1}(\tau))
        [\{\Phi^{-1}(\tau)\}^2-1]^2 \right]^{-1/5}
\end{align*}
Following \citet[eq.\ (3.30)]{Silverman1986}, to get a feasible bandwidth $\sthat{s}$, the unknown $\sigma$ is replaced by either the sample standard deviation of the $\sthat{v}_i$ or their sample interquartile range divided by $1.349$ (the standard normal interquartile range), whichever is smaller.
This $\sthat{s}$ is then used to compute $\sthat{f}_v(0)$ using \eqref{eqn:f0-hat}.

For the density derivative $f_v'(0)$ (here simply $f'(0)$), the approach is the same: use a nonparametric kernel estimator with a Silverman-type bandwidth.
The usual estimator is
\begin{equation}
\label{eqn:fd0-hat}
\frac{1}{nb^2}\sum_{i=1}^{n}K'(-\sthat{v}_i/b)
\end{equation}
where now $b$ is the bandwidth (to avoid confusion with $h$ and $s$).
The asymptotic mean squared error for this estimator at the point zero is from \citet{WandJones1994},%
\footnote{Unfortunately, their (2.34) in section 2.12 has a typo, but the correct version of (2.34) can be derived from the bias and variance expressions in their exercise 2.6(a) on pages 52--53.}
\begin{equation*}
 n^{-1}b^{-3}f(0)
 \int_\R\{K'(v)\}^2\diff{v}
+(1/4)b^4\{f'''(0)\}^2
 \left\{\int_\R K(v)v^2\diff{v}\right\}^2
\end{equation*}
where $\int_\R K(v)v^2\diff{v}=1$ as in \eqref{eqn:mu2}.
Solving the first-order condition (setting the derivative with respect to $b$ equal to zero), this leads to the optimal bandwidth
\begin{equation*}
b^* = n^{-1/7} \left( \frac{3f(0)\int_\R\{K'(v)\}^2\diff{v}}{\{f'''(0)\}^2} \right)^{1/7} 
\end{equation*}
Plugging in the standard normal density for $K(\cdot)$,
\begin{equation*}
\int_\R\{K'(v)\}^2\diff{v}
= 1/(4\sqrt{\pi}) 
\end{equation*}
Plugging in $f(0)$ from \eqref{eqn:f0} and%
\footnote{\texttt{D[PDF[NormalDistribution[\textbackslash[Mu], \textbackslash[Sigma]], x], \{x,3\}]} at \url{http://www.wolframalpha.com} for example; then plug in $x=0$ and $-\mu/\sigma=\Phi^{-1}(\tau)$.}
\begin{equation*}
f'''(0)
= \sigma^{-3}f(0)\Phi^{-1}(\tau)
  \left[3 - \{\Phi^{-1}(\tau)\}^2 \right]
\end{equation*}
yields
\begin{align*}
b^* &=
n^{-1/7}
\left( \frac{3 f(0) / (4\sqrt{\pi})}%
            {\sigma^{-6}\{f(0)\}^2\{\Phi^{-1}(\tau)\}^2
  \left[3 - \{\Phi^{-1}(\tau)\}^2 \right]^2} \right)^{1/7}
\\&=
n^{-1/7}
\left( \frac{3 / (4\sqrt{\pi})}%
            {\sigma^{-6} \sigma^{-1} \phi(\Phi^{-1}(\tau)) \{\Phi^{-1}(\tau)\}^2
  \left[3 - \{\Phi^{-1}(\tau)\}^2 \right]^2} \right)^{1/7}
\\&=
n^{-1/7} \sigma
\left( \frac{0.423}%
            {\phi(\Phi^{-1}(\tau)) \{\Phi^{-1}(\tau)\}^2
  \left[3 - \{\Phi^{-1}(\tau)\}^2 \right]^2} \right)^{1/7}
\end{align*}
Again, to get a feasible bandwidth $\sthat{b}$, $\sigma$ is replaced by the sample standard deviation or interquartile range divided by $1.349$, whichever is smaller.
This $\sthat{b}$ is then used to compute $\sthat{f}_v'(0)$ using \eqref{eqn:fd0-hat}.

Finally, $\sthat{f}_v(0)$ and $\sthat{f}_v'(0)$ are plugged into the expression for $h^*$ to get the feasible plug-in bandwidth $\sthat{h}$.

\subsubsection{Gaussian reference bandwidth}

Here, the Gaussian reference approach is used directly for $h^*$, to simplify the density and density derivative of $v$.

From \eqref{eqn:f0}, assuming normality of $v$ yields $f(0)=\sigma^{-1}\phi(\Phi^{-1}(\tau))$.

Assuming normality, the first derivative of the density of $v$ evaluated at zero is%
\footnote{\texttt{D[PDF[NormalDistribution[\textbackslash[Mu], \textbackslash[Sigma]], x], \{x,1\}]} at \url{http://www.wolframalpha.com} for example; then plug in $x=0$ and $-\mu/\sigma=\Phi^{-1}(\tau)$.}
\begin{align*}
f'(0) &= 
-\sigma^{-1}(-\mu/\sigma)f(0) \\
f(0)/\{f'(0)\}^2
&= \frac{1}{\sigma^{-2}\{\Phi^{-1}(\tau)\}^2 \sigma^{-1}\phi(\Phi^{-1}(\tau))}
 = \frac{\sigma^3}{\{\Phi^{-1}(\tau)\}^2 \phi(\Phi^{-1}(\tau))}
\end{align*}
Plugging this into the expression for $h^*$ yields
\begin{align*}
h^* =
n^{-1/3} \sigma
\left(\frac{3d}{\{\Phi^{-1}(\tau)\}^2 \phi(\Phi^{-1}(\tau))}\right)^{1/3}
\end{align*}
and as usual $\sigma$ is replaced by the smaller of the sample standard deviation of the $\sthat{v}_i$ or the sample interquartile range of the $\sthat{v}_i$ divided by $1.349$.

\subsection{Standard errors}
\label{sec:methodology-SE}

By default or with \stcmd{reps(0)}, analytic standard errors are computed as follows; otherwise, bootstrap is used as in section \ref{sec:methodology-BS}.
The formulas here are special cases of (3.11) of \citet{ChernozhukovHansen2006}; (18), (23), and Theorem 5 of \citet{deCastroGalvaoKaplanLiu2019}; or Theorem 7 of \citet{KaplanSun2017}.
For notational simplicity, I leave the dependence on $\tau$ implicit.
Define an additive error term
\begin{equation}
\epsilon \equiv y - \vecf{x}'\vecf{\beta} .
\end{equation}
The first-order asymptotic distribution of the \stcmd{sivqr} estimator is
\begin{equation}
\sqrt{n}(\sthat{\vecf{\beta}} - \vecf{\beta})
\dconv
\NormDist( \vecf{0} , (\matf{J}'\matf{S}^{-1}\matf{J})^{-1} )
,\;
\matf{J} \equiv \E[ f_{\epsilon|\vecf{z},\vecf{x}}(0\mid\vecf{z},\vecf{x}) \vecf{z}\vecf{x}' ]
,\;
\matf{S} \equiv \tau(1-\tau)\E(\vecf{z}\vecf{z}')
,
\end{equation}
where $f_{\epsilon|Z,X}(0\mid\vecf{z},\vecf{x})$ is the conditional PDF of $\epsilon$ given $(\vecf{z},\vecf{x})$ evaluated at $\epsilon=0$.
The expression $\matf{J}^{-1}\matf{S}(\matf{J}')^{-1}$ is mathematically equivalent but requires inverting the asymmetric matrix $\matf{J}$ by itself.
The estimated asymptotic covariance matrix is
\begin{equation}
(\sthat{\matf{J}}{}'\sthat{\matf{S}}{}^{-1}\sthat{\matf{J}})^{-1}
,\quad
\sthat{\matf{S}} = \tau(1-\tau) \frac{1}{n} \sum_{i=1}^{n} \vecf{z}_i \vecf{z}_i'
,\quad
\sthat{\matf{J}} = \frac{1}{nh} \sum_{i=1}^{n} \phi(\frac{\vecf{y}_i-\vecf{x}_i'\sthat{\vecf{\beta}}}{h}) \vecf{z}_i \vecf{x}_i' ,
\end{equation}
using a conventional kernel density estimator for $\sthat{\matf{J}}$ with a Gaussian kernel $\phi(\cdot)$ and bandwidth $h$ being Silverman's rule of thumb \citep[\S3.4.2]{Silverman1986}, specifically $1.06n^{-1/5}$ times the minimum of either the sample standard deviation of the $\vecf{y}_i-\vecf{x}_i'\sthat{\vecf{\beta}}$ or their sample interquartile range divided by $1.349$.
Alternatively, the bandwidth in (A.8) of \citet{Liu2019} would also work well.
Note $\sthat{\matf{J}}$ is a \citet{Powell1986} type estimator, similar to (3.14) of \citet{ChernozhukovHansen2006} but here with a Gaussian instead of uniform kernel.
The reported standard errors are then the square roots of the diagonals of the matrix $(\sthat{\matf{J}}{}'\sthat{\matf{S}}{}^{-1}\sthat{\matf{J}})^{-1}/n$.

\subsection{Bayesian bootstrap}
\label{sec:methodology-BS}

When \stcmd{reps(\num)} specifies multiple bootstrap replications, standard errors are computed by Bayesian bootstrap \citep{Rubin1981}, assuming i.i.d.\ sampling.
(With non-i.i.d.\ sampling, \stcmd{bootstrap} can be used; see \rref{bootstrap}.)
This is a particular type of frequentist exchangeable bootstrap \citep[for example,][ex.\ 3.6.9, p.\ 354]{vanderVaartWellner1996} that also has a nonparametric Bayesian interpretation \citep[for example,][who advocate for Bayesian bootstrap and show IV and QR applications]{ChamberlainImbens2003}.
In each replication, weights are set as $w_i=\xi_i/\stbar{\xi}$ where the $\xi_i$ are i.i.d.\ standard exponential random variables and $\stbar{\xi}$ is their average; equivalently, the vector $(w_1,\ldots,w_n)/n$ follows a Dirichlet distribution with every parameter equal to one.
Using these $w_i$, the estimator is computed by solving \eqref{eqn:SEEw}.
This is repeated many times; the reported standard error is the standard deviation of the different $\sthat{\vecf{\beta}}$.

\section{Conclusion}
\label{sec:conc}

The \stcmd{sivqr} command offers smoothed estimation of the instrumental variables quantile regression model of \citet{ChernozhukovHansen2005}.
The smoothing improves computation and accuracy, helping \stcmd{sivqr} to overcome several limitations of \stcmd{ivqreg} and \stcmd{ivqreg2}, while complementing commands for alternative quantile models with endogeneity (\stcmd{cqiv} and \stcmd{ivqte}).
The \stcmd{sivqr} command can help \Stata\ users reliably estimate heterogeneous effects across a variety of settings and datasets.

\section{Acknowledgements}
\label{sec:thx}

There would be no \stcmd{sivqr} \Stata\ command without the work of Yixiao Sun on \citet{KaplanSun2017}.
Thanks to Xin Liu for research assistance.
Thanks to Joao Santos Silva for suggesting the \stcmd{initial()} option.
Thanks to Di Liu for several helpful suggestions and encouragement, including \stcmd{e(bwidth\_max)} and analytic standard errors.
Many thanks to a particularly thorough anonymous reviewer who found my mistakes and helped improve both the content and presentation tremendously.



\bibliographystyle{sj}
\ifnum 35=1 \def\bibname{Reference}
\else \def\bibname{References} \fi


\begin{aboutauthor}
David M.\ Kaplan is an associate professor in the Department of Economics at the University of Missouri.
His primary research interest is econometric methodology.
In particular, he enjoys creating and advancing methods for understanding changes and treatment effects on entire distributions (instead of just averages), also a focus of his free open-source textbook and his first \Stata\ command, \stcmd{distcomp}.
\end{aboutauthor}

\clearpage
\end{document}